\documentclass[a4paper,fleqn]{cas-sc}

\usepackage[authoryear,longnamesfirst]{natbib}
\usepackage{subcaption}

\def\tsc#1{\csdef{#1}{\textsc{\lowercase{#1}}\xspace}}
\tsc{WGM}
\tsc{QE}
\tsc{EP}
\tsc{PMS}
\tsc{BEC}
\tsc{DE}

\begin{document}
\let\WriteBookmarks\relax
\def\floatpagepagefraction{1}
\def\textpagefraction{.001}

\shorttitle{Behavioural ride-pooling}

\shortauthors{Bujak, M. and Kucharski, R.}

\title [mode = title]{Ride-pooling service assessment with rational, heterogeneous, non-deterministic travellers}                      

%
\author[1,2]{Michal Bujak}[type=editor,
                        orcid=0000-0001-5383-6459]

\cormark[1]


\ead{michal.bujak@doctoral.uj.edu.pl}



\affiliation[1]{organization={Department of Mathematics and Computer Science, Jagiellonian University},
    addressline={ul. Prof. Lojasiewicza 6}, 
    city={Krakow},
    postcode={30-348}, 
    country={Poland}}

\author[1]{Rafal Kucharski}[orcid=0000-0002-9767-8883]
\ead{rafal.kucharski@uj.edu.pl}
\ead[URL]{https://rafalkucharskipk.github.io/}

\affiliation[2]{organization={Doctoral School of Exact and Natural Sciences, Jagiellonian University},
    city={Krakow},
    postcode={31-007}, 
    country={Poland}}

\cortext[cor1]{Corresponding author}



\begin{abstract}
Ride-pooling remains a promising emerging mode with a potential to contribute towards urban sustainability and emission reductions. 
Recent studies revealed complexity and diversity among travellers' ride-pooling aptitudes. 
So far, ride-poling analyses assumed homogeneity and/or determinism of ride-pooling travellers.
This, as we demonstrate, leads to a false assessment of ride-pooling system performance. 
We experiment with an actual NYC demand from 2016 and classify travellers into four groups of various ride-pooling behaviours (value of time and penalty for sharing), as reported in the recent SP study.
We replicate their random behaviour to obtain meaningful distributions.

\noindent Unsurprisingly, results vary significantly from the deterministic benchmark: expected mileage savings were lower, while the utility gains for travellers were
greater.
Observing performance of heterogeneous travellers, we find that those with a low value of time are most beneficial travellers in the pooling system, while those
with an average penalty for sharing benefit the most.
Notably, despite the highly variable travellers' behaviour, the confidence intervals for the key performance indicators are reasonably narrow and system-wide performance remains predictable.

\noindent
Such findings shed a new light on the expected performance of large scale ride-pooling systems. We argue, that the policy recommendations shall be revised to accommodate behavioural heterogeneity.
\end{abstract}


\begin{highlights}
\item We assess the excepted ride-pooling system performance assuming the heterogeneous behaviour observed in recent studies.
\item We consider four groups of travellers varying with value of time and willingness to share, and replicate the probabilistic ride-pooling problem to obtain meaningful distributions.
\item Results differed significantly from the deterministic benchmark: expected mileage savings were lower, while the utility gains for travellers were greater.
\item Ride-pooling systems' implementation shall take into account heterogeneous groups of travellers.
\item System performance for low value-of-time travellers is substantially different from the one used by high value-of-time travellers.
\end{highlights}


\maketitle

\section{Introduction}

\begin{figure}[!h]
    \begin{center}
      \includegraphics[width=\textwidth]{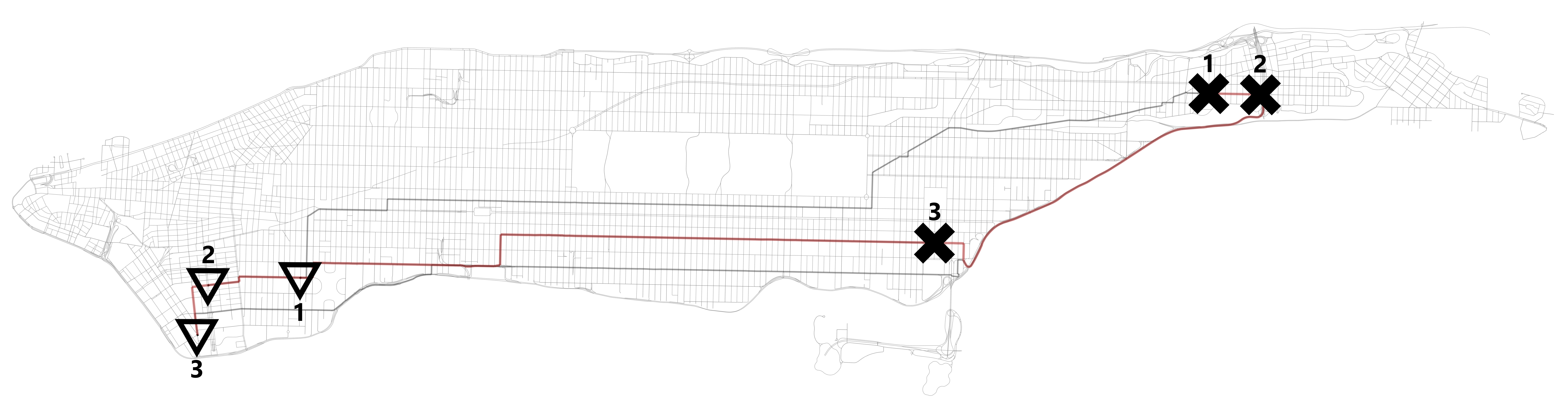}
    \caption{\small Ride-pooling illustrated. Three travellers request to travel from their origins (marked with $\times$) to destinations (marked with a $\mathcal{r}$). Apart from offering them three separate solo-rides (marked with grey), platform offers to pool them together and serve with a single trip (red). This induces a discomfort (sharing trip with others), detours (route is longer) and delay (three travellers need to coordinate their departure times) which is compensated with a discounted trip fare.  Will the travellers find the pooled option attractive? This depends on their behavioural profiles. So far travellers were assumed either captive, homogenous and/or deterministic. Here, we utilise the recent findings on heterogeneous, non-deterministic ride-pooing behaviours. Each of three travellers may now have different value-of-time or willingness to share and decide to opt-out from the pool. We investigate how such heterogeneity impacts the expected system performance.  } 
    \label{fig:illust}
    \end{center}
\end{figure}

Ride-pooling is a ride-hailing service which enables travellers to share a ride. Two or more trip requests submitted to the ride-hailing platform (like Uber, DiDi or Lyft) are pooled into a single vehicle, which picks consecutive travellers at their origins and takes them to their respective destinations, as illustrated in Fig. \ref{fig:illust}. Such pooled ride is typically longer (due to pick-up delay and detour) and yields a sharing discomfort, which are compensated by a lower ride fare. 

Ride-pooling, despite being a promising emerging mode of urban mobility, still struggles to gain momentum. Pre-pandemic it managed to reach critical mass and become profitable for providers and attractive for travellers (\cite{shaheen2016shared}). Yet, in the post-pandemic landscape of urban mobility, it needs to grow again. 
Ride-pooling services, when deployed successfully and on a bigger scale, can substantially contribute towards sustainability transitions. Offering an attractive alternative to private car usage or solo ride-hailing, yet with greater occupancy. When designed carefully, it can complement the public transport offering attractive first-last mile services, or being an alternative in the areas and time-periods where high frequency mass transit is unnecessary. This is challenging, specifically when the perspective of four different parties need to be aligned: travellers (who want to arrive on time, with high comfort and at the low price), drivers (who want to maximise their earnings), platform (who wants to gain market share and benefit from high commission rates) and the city (who wants to reach the objectives of its transport policy). Such alignment is not trivial, specifically when we acknowledge that individual agents are heterogeneous and have different behavioural traits, leading to different perceptions of the new service. Here, we contribute to that alignment and demonstrate how can we assess the ride-pooling system's performance with heterogeneous travellers.

While the early studies on ride-pooling focused on the computational complexity and challenges of real-time operations, the perspective of travellers was often neglected. State-of-the-practice algorithms  \cite{alonso2017demand} successfully match travellers into high order groups, served in real-time with ride-hailing vehicles. Though they often treat travellers captive, insensitive to delay and detour of the ride-pooling services as long as they fit into arbitrary time-windows. In reality, travellers are rational utility maximisers, extremely sensitive to the service performance. Naturally opting-out from the systems which they perceive not optimal. In that notion, we built our previous ride-pooling algorithm ExMAS \cite{kucharski2020exmas}, strictly limiting the ride-pooling to the rides attractive for the travellers, for which the utility is greater than for the alternative (e.g. private ride-hailing ride). 

Here, we move further along this line and introduce the natural heterogeneity into the picture. For new mobility modes, the perceptions may differ dramatically, from enthusiasts to sceptics. This is no different in ride-pooling where some travellers may avoid crowding, be afraid of infectious diseases, aim to save time or save money. Unsurprisingly, such heterogeneity was observed in empirical studies \cite{alonso2021determinants}, but hitherto not implemented into algorithms and experimental studies. As we demonstrate, neglecting the fact that the decision to ride-pool is non-deterministic and made by heterogeneous agents with substantially different attitudes, leads to false assessments. 

Ride-pooling needs a critical mass to gain momentum and take-off, which leads to non-linearity in the system performance. However, not only the total number of clients in the system, but also the presence and properties of some individual travellers may be crucial. Thus, the assessment based on the mean, average expected behaviour of ride-pooling individuals is likely to be not representative for the performance of the whole system. When we replace the average shareability of travellers (behavioural preferences regarding pooling) with individual properties, we will observe on one hand isolated travellers unwilling to share with anyone and on the other extremely shareable individuals happy to join any other pooled group. This has a highly asymmetrical impact on the structure of the underlying shareability graph and thus on the ride-pooling performance, we observe it e.g. through the maximal size of pooled rides, which increases when travellers become non-deterministic.

 Using findings from an empirical study \cite{alonso2021determinants}, which revealed four distinct classes of ride-pooling travellers and observed variability in their ride-pooling decisions, we introduce two levels of non-determinism into the ride-pooling modelling. First, we assume distinct attitudes for ride-pooling, i.e. latent classes associated with different perceptions and valuations of crucial ride-pooling service variables. Second, we introduce an error term to the ride-pooling perceptions made by individual travellers. 

In the experimental part, we take the side of the operator which wants to estimate the potential of introducing ride-pooling. We assume that the operator can predict the demand for her services in terms of trip requests (origin, destination and time), but she is not aware of the specific attitudes of particular clients. She does not know their exact value-of-time and willingness-to-share. Though she knows the system-wide composition of the population (from empirical studies, like \cite{alonso2021determinants}) and can decompose the population into distinct groups of more definite behavioural traits.

Namely, we can tell that each of our clients belongs to a given behavioural class (group) with a given probability and this class is associated with given ride-pooling behaviour. This additional knowledge comes at the cost of non-determinism, which we need to incorporate into our assessment. The final ride-pooling decision of each individual is unknown for the provider and depends on the class membership (which is random) and the value of the error term (which is random again). To obtain a meaningful ride-pooling picture in such probabilistic setting, we replicate this random process sufficiently enough to estimate the distributions, which are not evident, as we report.

We run the experiment in New York, where we use the actual travel demand observed for taxi trips. We use a 30-minute batch of $150$ trip requests and match travellers into attractive pooled rides. We assume behavioural parameters as they were revealed in the empirical stated-preference study by \cite{alonso2021determinants}, in particular the values of value-of-time and willingness-to-share. After running 1000 replications of underling random process, we obtain results that significantly enrich the deterministic picture.

Such a perspective allows the provider for more meaningful estimation of the system performance and its distribution. Our results, apart from the expected values, reveal confidence intervals and tails. Estimates of performance (e.g. vehicle kilometres travelled, fares collected) and perceived attractiveness for travellers are now more meaningful and robust. Policymakers and system operators may now understand better the different points of view and tailor the system to the needs of varying target groups. Specifically important for policymakers who consider subsidising ride-pooling services. With our method, they can distinguish when the service is used primarily as the cheaper taxi (and shall not be publicly subsidised in general) from the desired cases when it complements the public transport and aligns with policy objectives. The system deployed to serve the groups targeted by the policymakers (e.g. groups with lower accessibility and equity) will have different performance and can be tailored accordingly. Covering the specific needs of the target groups increases the chances that the system reaches stability and sustainability.

\subsection{Related work}

Ride-pooling is environmental- and city-friendly service compared to standard ride-hailing services offered by transport network companies (TNCs). \cite{fagnant2014travel} argues that one shared vehicle can replace around eleven privately owned vehicles. Benefits of partial adoption of on-demand vehicles for ride-pooling service are discussed in \cite{engelhardt2019quantifying}. \cite{martinez2015agent} proposes a ride-pooling system which, while added to the standard ride-hailing system, is beneficial for users. \cite{zhang2021pool} discusses the up-sides and down-sides from travellers’ and operator’s perspective. \cite{martinez2017assesing} provides an insight into CO2 reduction. The most common measure for environmental benefits is vehicle kilometres reduction, widely reported in the studies, like \cite{bilali2020analytical} which provides both theoretical estimation and simulation results of ride-pooling efficiency in terms of reduced vehicle hours. Researchers document different values (dependent upon demand and city structure, matching algorithm): \cite{zhu2022potential} finds the total vehicle distance reduced by only $8.21\%$, \cite{chen2017data} provides the value of $16\%$, while the seminar paper by \cite{santi2014quantifying} suggests savings exceeding $30\%$. 

Ride-pooling can be analysed on many levels. Ramifications of spatiotemporal demand structure are analysed for two major German cities by \cite{zwick2022ride}. \cite{soza2022shareability} analyses how ride-pooling performance changes with the various synthetic demand distributions. Operational challenges on the supply side are analysed in the paper by \cite{zwick2022shifts}. \cite{zhang2021pool} analyses pricing, finds market equilibrium and proposes congestion tax policy. Competition in the duopoly ride-sourcing market is analysed by \cite{guo2022day}. \cite{li2022optimal} analyses optimal fare and fleet size with congestion effects.

Variety of algorithms were proposed to solve the challenging ride-pooling problem. In the seminal work, \cite{santi2014quantifying} introduces a real-time ride-pooling algorithm and invaluable shareability networks, however limits his solution to two sharing travellers. \cite{alonso2017demand} proposed a state-of-the-art real-time algorithm which allows a higher-degree rides and is considered the benchmark for other approaches. Above authors, like many others (\cite{ke2021probability}, \cite{shah2020neural}, \cite{bilali2020analytical}, \cite{wang2021predicting}) use fixed constraints, typically on maximal pick up delay and detour time. The fixed constraints are irrespective of total travel time duration and individual preference. 
An alternative approach is proposed by \cite{kucharski2020exmas}. The proposed ExMAS algorithm (detailed in the Methodology section) is an offline algorithm which has relative constraints posed by travellers. The acceptable pick-up and detour time is determined by unified behavioural parameters (such as value of time, penalty for sharing) and travellers are pooling only if it is attractive for them.

To determine performance of the ride-pooling system, \cite{ke2021probability} analyses three aspects: fraction of passengers who effectively share a ride, average detour of a traveller and average vehicle routing distance. \cite{alabassi2019deeppool} introduces a deep neural network with the objective function aiming to minimise four objectives: supply-demand mismatch, waiting time and detour distance, travel time extension and number of vehicles and resources used. \cite{agatz2012optimization} lists three system-wide performance indicators: vehicle distance, travel time, fraction of pooled passengers. 

Many studies have been conducted to understand ride-pooling preferences (\cite{krueger2016preferences}, \cite{alonso2021determinants}, \cite{lavieri2019modeling}). \cite{chavis2017development} examines preferences towards three services: fixed, flexible and individual, and estimates alternative specific constants of route type along with explanatory (travel time, cost, etc.), socioeconomic and attitudinal variables. \cite{krueger2016preferences} introduces three alternatives: non-shared ride-hailing, ride-pooling and public transit. The primary variable is the value of time, which is further split into in-vehicle time and waiting time. In a stated preference survey \cite{gervzinivc2022potential} assesses the demand for on-demand and pooled mobility services in urban areas. \cite{alonso2021determinants} investigates in depth ride-pooling preferences in ride-hailing services. They design a stated preference study and (based on the sample of 1000 Dutch respondents) disentangle the sharing aspect from related time–cost trade-offs (e.g. detours), investigate preference heterogeneity and identify distinct market segments.
The study, which results we apply in our experiments, identifies four classes with distinct behavioural characteristics, including value of time and willingness to share (penalty for sharing).

\subsection{Study objectives and approach}
So far, the utility-based approaches for ride-pooling are scarce, the focus is on real-time operations, assuming travellers are captive. In the rare approaches, where the traveller is assumed to be a rational utility maximiser, the decision process is simplified. Travellers are assumed deterministic and/or homogeneous, which contradicts recent empirical observations in which a strong heterogeneity and non-determinism of ride-pooling travellers was revealed. 

In this study, we fill this gap. We propose an off-line approach where we assume the demand is known in advance. We pool travellers into attractive shared rides and find the optimal matching (minimising the total mileage). Unlike the real-time approaches, we do not rely on fixed time-window constrains to determine if the pooling is attractive, but we explicitly treat travellers as decision makers and reproduce their process of opting for the subjectively optimal service.

While such approach better fits the reality, it is also challenging to reproduce and to interpret the results. Adding two levels of noise into the choice process makes the whole analysis less straightforward. First, the practical experimental setting needs to be well-designed to represent the actual and practical problem. Second, the underlying stochastic process needs to be replicated to obtain meaningful results, which calls for rigorous and non-trivial interpretation.

Eventually, by incorporating the notion of non-determinism and heterogeneity into the ride-pooling we may address the following problem: What is the expected performance of ride-pooling system for which we know the detailed spatio-temporal demand pattern and the behavioural structure of the population, yet the exact behaviour of  individual clients remains latent? Such problem statement setting, while clearly impractical for real-time operations, significantly improves strategic assessment by providing meaningful estimates of vehicle hours reductions, perceived attractiveness or profitability. 

Our findings are instrumental for policymakers and platform decision makers who want to enrich their strategic decisions with the notion of travellers' heterogeneity. Policymakers may understand when and where the ride-pooling system remains efficient, i.e. positively contribute towards sustainability transitions. Platform executives may use it to make optimal business decisions, which, in turn, make the ride-pooling operations profitable. Based on such analysis, the platform provider may reasonably identify e.g. the optimal discount rate at which ride-pooling service, for the expected demand, is both reliable, attractive and profitable. This is particularly timely post-pandemic, when reintroduced, ride-pooling services offer significantly lower discounts for ride pooling than pre-pandemic (like Uber: comp. \cite{lo2018perfect} and \cite{UberShare}). Other platform providers may want to identify what kind of offer would be their best response. And this is crucial not only for the benefits of a particular provider, but also system-wide - which we want to benefit from ride-pooling.

\section{Methodology}\label{sec:Methodology}

To reach the objectives of this study, we extend the previously proposed off-line utility-based ride-pooling algorithm ExMAS. We introduce non-deterministic utility formulas to integrate them with the previous deterministic approach and solve the ride-pooling problem for heterogeneous population. ExMAS solves the ride-pooling problem, i.e. for a given demand (set of trip requests) it first identifies the attractive pooled groups (rides) of any size (degree) and then optimally assigns each traveller to a particular ride (pooled or private). 

In the non-deterministic case, we first sample the behavioural profiles of individual travellers, then we identify the attractive pooled groups and identify matching that minimises the total mileage. From such solution we estimate the central KPIs of the system: mileage reductions, perceived attractiveness, passenger detour and profitability. In the experiments, we replicate this process sufficiently enough to obtain meaningful distributions of random variables (KPIs).

\subsection{Ride-pooling problem solution with ExMAS}\label{sec:exmas}
Exact matching of attractive shared rides (ExMAS, \cite{kucharski2020exmas}) is a strategic, utility-driven ride-pooling algorithm. In contrast to e.g. \cite{alonso2017demand}, ExMAS works offline, assuming a complete knowledge of the demand and (since it takes the strictly demand-oriented perspective) does not explicit consider individual vehicles. While such an approach is clearly not suitable for real-time operations, it is invaluable for strategic, demand-oriented analyses. Instead of fixed time windows or spatial constrains in which travellers are pooled regardless of their preferences,  the focus is on the passengers' utility.

More strictly, two or more passengers are assigned to a shared ride if the pooled ride utility is greater than a solo-ride utility for each of them. Utility is generically composed of travel times, costs and their relative weights (behavioural parameters). In the original ExMAS, the (dis)utility of a shared- (for passenger $i$ participating in ride $r_k$ denoted as $U^s_{i, r_k}$) and of the non-shared ride (denoted $U^{ns}_i$) were the following: 

\begin{align}\label{utility_original_exmas}
\begin{split}
& U^{ns}_i = -\rho l_i - \beta_t t_i \\
& U^s_{i, r_k} = -(1 - \lambda) \rho l_i - \beta_t \beta_s (\hat{t}_i + \beta_d(\hat{t}^p_i)) + \epsilon,
\end{split}
\end{align}

where $\rho$ stands for per-kilometre fare, while $\lambda$ denotes discount for sharing a ride. Both are controlled by the operator. $\beta^t$, $\beta^s$ $\beta^d$ are the behavioural parameters: value of time, penalty for sharing and delay sensitivity, respectively. $t_i$ and $\hat{t}_i$ stand for travel time of non-shared and shared ride, respectively, $\hat{t}^p_i$ is a pick-up delay. Travel time with the shared ride is updated for each evaluated ride candidate and is typically greater than $t_i$ due to both pooling detour (to pick-up and drop of other travellers) and delay (to wait for others). $\epsilon$ is a random term.
Hitherto, all behavioural parameters ($\beta$'s) were assumed homogeneous among the population and the problem was simplified by the assumption that $\epsilon = 0$ which makes the problem purely deterministic. Consequently, a shared ride $r_k$ was attractive by the $i$-th traveller if simply $U^s_{i, r_k} > U^{ns}_i$.

ExMAS uses the above formulation to evaluate if any theoretically feasible group of travellers is attractive. Starting with pairs of travellers and then gradually increasing the degree (size) of the identified pooled rides. Sharing discount is assumed to be flat, regardless of number of co-travellers, hence attractive ride of higher order comprises attractive rides of lower order. Applying the recursion and exploiting the shareability graph properties, ExMAS identifies for potential attractive triples by combining pairs, later quadruple based on triples, etc. The search is exhaustive, yet the filtering for attractive rides only, prevents the search space explosion in practical settings.

Identified rides may be represented with a so-called shareability networks, one of the outcomes of ExMAS that we exploit in this study. Two networks are generated: in the \textit{shareability network} two nodes (travellers) are linked when they can be attractively pooled, while in the \textit{matching network} two nodes (travellers) are linked if they actually travel together in the final solution (bi-partite matching), as we illustrate in the results.

From the set of all feasible and attractive pooled rides we may identify the optimal solution. Solution is found with a bi-partite \textit{matching} problem minimising the total vehicle hours with a constraint that each passenger is served with exactly one ride (pooled or private). 
The output is a set of rides with their schedules (subsequent pick-up and drop-off points and times served along the ride) from which we can measure the ride-pooling system’s performance.

Full details on the algorithmic steps in ExMAS are described in \cite{kucharski2020exmas}. Notably, ExMAS is demand-driven and thus the vehicles (fleet) is not considered explicitly with a reasonable assumptions that: a) the fleet size needed to serve pooled services is not greater than serving individual private rides and b) when the degree of attractive pooled rides starts exceeding the capacity of a single vehicle (three travellers) the service reached a critical mass advocating for bigger vehicles (e.g. 8-person vans).

\subsection{Ride-pooling utility for non-deterministic heterogeneous travellers}\label{sec:methodology_behavioural}
In the study, we extend our previous approach to account for the heterogeneous behavioural characteristics of individuals. Namely, any of behavioural parameters ($\beta$'s in eq. \ref{utility_original_exmas}) instead of being constants, can be now a random variable. 
We assume that preferences of an individual are unknown, while we know the distributions within the population.
Moreover, to account for additional uncertainty, we introduce a panel noise $\epsilon_i$ (traveller-specific) and noise $\epsilon_{i, r}$ (ride-specific for specific traveller), leading to the following formulas:

\begin{align}\label{utility_new_exmas}
& U^{ns}_i = \rho l_i + \mathring{\beta_t} t_i \\
& U^s_{i, r_k} = (1 - \lambda)\rho l_i + \mathring{\beta_t} \mathring{\beta_s} (\hat{t}_i + \mathring{\beta_d}(\hat{t}^p_i)) + \epsilon_i + \epsilon_{i,r},
\end{align}

where all notions introduced in the eq. \ref{utility_original_exmas} still apply, yet $\mathring{\beta}$s are now random variables.  Consistently with a discrete choice theory, we assume that the probability that a traveller $i$ finds a shared ride $r_k$ attractive is expressed with: 
\begin{equation}
P^s_{i, r_k} = \Pr ( U^s_{i, r_k} > U^{ns}_i),
\end{equation}
which depending on the random variables distributions may become e.g. a logit or a probit model.
While in general any distribution of random variable can be applied in the method, in the experiment we use multiple classes $\mathcal{C}$ and assume the random variables to follow a multimodal normal distribution (unimodal within the class):

\begin{equation}
    \mathring{\beta_j} = \sum_{C \in \mathcal{C}} p_{C} X(\bar{\beta}_{j,C},\sigma_{j,C}) \mathrm{\ for\ } j \in \{c, t, s,  d \},  \label{eq:random}
\end{equation}

where $p_{C}$ is the probability of belonging to the class $C$, $X(\bar{\beta}_{t,C},\sigma_{j,C})$ is a random variable following normal distribution with mean of $\bar{\beta}_{t,C}$ and a standard deviation of $\sigma_{j,C}$ (coefficient- and class-specific). 

The above formula may be applied for any behavioural parameter, yet in our experiment only $\beta_t$ and $\beta_s$ are random variables. Which is in-line with findings from recent studies on behavioural heterogeneity among ride-pooling travellers' (\cite{krueger2016preferences}, \cite{alonso2021determinants}, \cite{lavieri2019modeling}). Value of time serves as a proxy between longer travel time and monetary savings. Penalty for sharing reflects an individual’s attitude towards sharing (including infection awareness), which varies substantially. While VoT and PfS are collinear in $U^s$ formula, VoT is the only one to impact $U^{ns}$, which disentangles the impact of the two variables.

Mind that the two noise terms in eq. \ref{utility_new_exmas} are applied differently, the first-one $\epsilon_i$ is the traveller-specific and is drawn once for each ride-pooling problem.
$\epsilon_i$ can be interpreted as an attitude towards pooling, irrespective of the trip’s characteristics. While the traveller remains a rational decision maker, there is always uncertainty associated with choice which is not captured by the utility formula, expressed with $\epsilon_{i,r}$ (drawn at each evaluation of potentially attractive pooled ride). 

On the final note, we assume that class-membership probability $p_C$ is an independent random variable, yet this can be easily enhanced e.g. with sociodemographic variables (when specific age, gender or income groups constitute respective classes). In the practical scenario, the platform may either successfully learn or predict the class membership of the specific $i$-th traveller $\mathbf{P}({C} \text{\textbar} i)$, effectively improving estimation. Unfortunately, such data was not included neither in the demand set that we used, nor in the choice model estimation of \cite{alonso2021determinants}.

Subsequent to utility determination, the remainder of the ride-pooling algorithm remains the same as summarised in Section \ref{sec:exmas} and detailed in the \cite{kucharski2020exmas}.

\subsection{Performance indicators}\label{sec:performance_indicators}
We measure the performance of ride-pooling solution with the four following indicators. For the environment and city perspective we look at the vehicle-kilometres saved due to pooling $\mathcal{D}$. For travellers, we observe pooling costs (relative detours $\mathcal{T}$) and benefits (relative improvement in utility $\mathcal{U}$). For the platform, we look at the profitability of ride-pooling observations $\mathcal{P}$.

$\mathcal{D}$ expresses the relative mileage savings as follows:

\begin{equation}\label{func:D}
    \mathcal{D} = \frac{D_{ns} - D_{s}}{D_{ns}},    
\end{equation}
where $D_{ns}$ denotes distance in the non-shared scenario (e.g. if everyone travelled alone) and $D_s$ - total lengths of rides comprising the ride-pooling solution.

Mind that here, we use only distance with travellers and do not include deadheading (distance travelled without travellers) - which we assume do not increase due to sharing (compared to private rides).

From the traveller's perspective, we measure two quantities: trip detour and utility gain. Similarly to $\mathcal{D}$, we introduce functions $\mathcal{T}$ and $\mathcal{U}$, given as:
\begin{align}
    \mathcal{T} = \frac{T_{s}- T_{ns}}{T_{ns}}, \label{func:T} \\
    \mathcal{U} = -\frac{U_{s} - U_{ns}}{U_{ns}}, \label{func:U}
\end{align}
where $T_{ns}$ and $U_{ns}$ denote total travel time and utility of all the travellers in non-shared scenario while $T_s$ and $U_s$ corresponding values in ride-pooling scenario. While formally the travel time in shared option is longer due to both detour (longer route due to pooling with others) and delay (waiting for others), here, for brevity, we simply call it a detour. Mind the negative sing in eq. \ref{func:U}, added since the utility (like usually in the transportation) is the disutility (composed of monetary costs and time) and is always negative.

We introduce the profitability $\mathcal{P}$, being greater than $1$ when the provider's mileage savings are greater than loss of revenues (due to sharing discount). 
First, we introduce it at the ride level:
\begin{equation}\label{func:P}
    \mathcal{P}_r = \begin{cases}
    \frac{(1-\lambda)\lambda_p \sum_{i \in r} d^{ns}_i}{\lambda_p d_r} =\frac{(1-\lambda) \sum_{i \in r} d^{ns}_i}{d_r}& \text{, if ride is pooled} \\
    1              & \text{, otherwise,}
\end{cases}
\end{equation}
where $d^{ns}_i$ denotes the direct trip length (traveller $i$ pays for the distance that would be served with the private ride, not with the detour), $d_r$ is a total mileage of the ride $r$ shared by travellers $i \in r$, $\lambda$ stands for sharing discount and $\lambda_p$ is fare. The second term reflects the travellers who were not pooled with anyone and paid the full price.

Profitability $\mathcal{P}$ on the system-wide level is then:
\begin{equation}
    \mathcal{P} = \frac{\sum_{r \in \mathbf{R}}  \mathcal{P}_r d_r}{\sum_{r \in \mathbf{R}} d_r},
\end{equation}
where $\sum_{i \in \mathbf{R}} d_r$ is a total length of all rides in the solution.

The above indicators can be calculated not only at the system-level (summing over all travellers), but also for individual travellers ($\mathcal{T}$ and $\mathcal{U}$) or individual rides ($\mathcal{D}$ and $\mathcal{P}$).
We will use subscript $r$ or $t$ for reference to individual rides' ($\mathcal{D}_r$, $\mathcal{P}_r$) or traveller's ($\mathcal{T}_t$, $\mathcal{U}_t$) performance.

\subsection{Assessment}
Introduction of randomness when determining whether a pooling option is attractive or not, results in a non-deterministic solution and thus a system’s performance. 
Importantly, there are two intuitive perspectives on which one can assess the variability. One is to look at the solution of a single ride-pooling problem and apart from identifying the mean performance (e.g. utility gains $\mathcal{U}$) we look at how this total is distributed across the individual travellers $\mathcal{U}_i$. In this study, however, we replicate the random process (due to non-deterministic behaviour) and see how the performance vary across the replications (as we report e.g. in Fig. \ref{fig:kpis}). Moreover, two approaches can be joined when we look at individual travellers' performance across the replications, like we report e.g. in Fig. \ref{fig:scatters}.

\section{Results}

We illustrate how the proposed method enhances assessment of ride-pooling services with the case-study of New York City. We reproduce the likely case when the service provider can predict the demand and its behavioural structure, yet the actual characteristics of individual behaviours remain latent. 

We start with parametrising the behaviour of four distinct traveller groups revealed in empirical study in Section \ref{sec:results_behavioural_component}. Then, we  visualise the underlying shareability networks and show how they evolve in the non-deterministic setting (Fig. \ref{fig:nets}). We show the distributions of key ride-pooling performance indicators in Fig. \ref{fig:kpis} and compare with the deterministic benchmarks, further stratified for four behavioural classes in Fig. \ref{fig:cdfs}. We report the asymmetrical impact of individual willingness-to-share on system performance and we show, in Fig. \ref{fig:rides_profitabilit}, how the maximal number of travellers riding together increases in a non-deterministic case. Then, we report how varying behaviour of individuals impacts their ride-pooling performance in Fig. \ref{fig:scatters}. We conclude with reporting how the variability of ride-pooling behaviour scales with the demand in Fig. \ref{fig:scaling}.

\subsection{Experiment details}
\begin{figure}
    \centering
    \includegraphics[width=.85\textwidth]{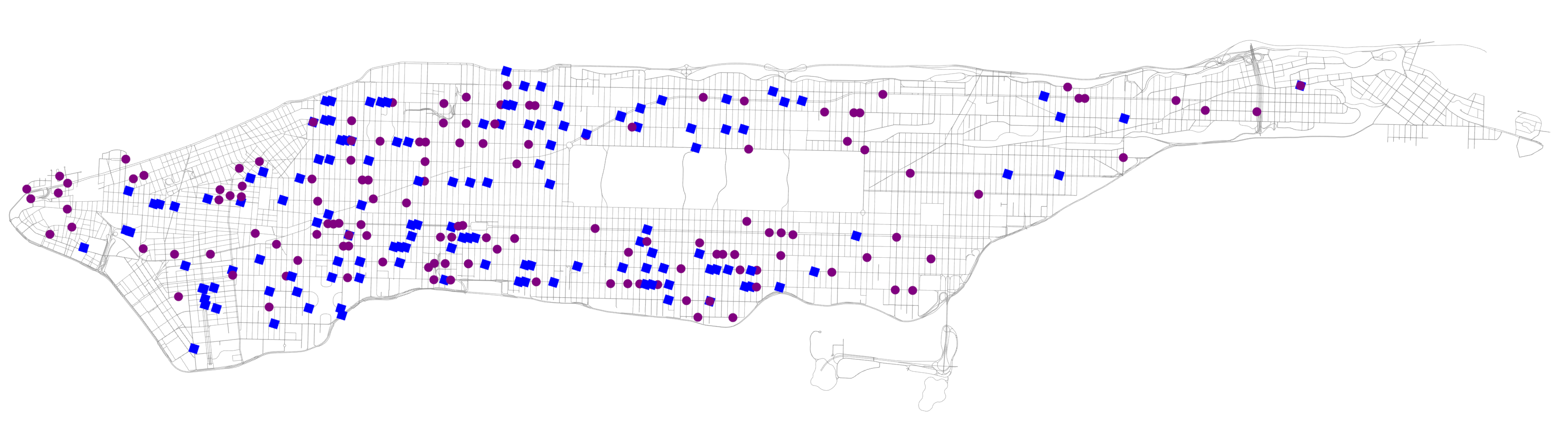}
    \caption{\small Demand dataset for experiments: ca. $150$ trip requests from Jan 2016 in Manhattan. Blue dots are origins, purple destinations.}
    \label{fig:demand}
\end{figure}

We experiment with the trips actually requested in the New York City on January $2016$. We use the set of about $150$ (exactly $147$) trips requested in the $30$ minute batch (which is beyond the critical mass needed to induce pooling). Each trip request links origin with destination at a given time (as illustrated in Figure \ref{fig:demand}). We used the fare $\lambda_p$ of $1.5$ \texteuro/km (converted from \$, consistent with \cite{taxi_fare_calculation}), sharing discount $\lambda$ is $30\%$ (within the range suggested by \cite{shaheen2019shared}). We assume $\beta_d = 1$ while the rest of behavioural parameters is detailed in the following section (\ref{sec:results_behavioural_component}).

We assume that the service provider can reasonably estimate the travel demand (trip requests) and knows the structure of travellers' behaviour, yet cannot predict which requests belong to which classes. 
To produce reliable estimates, we fix the demand (requests) and replicate the experiment (assuming various ride-pooling behaviour of individual travellers) for $1000$ times. Results are available to reproduce on the public repository\footnote{Script to reproduce results is on the branch \textit{probabilistic\_topological} of original ExMAS (direct link: \url{https://github.com/RafalKucharskiPK/ExMAS/blob/probabilistic_topological/Utils/Probabilistic_ExMAS_wrapper.py).}}.

\subsection{Behavioural component}\label{sec:results_behavioural_component}
Out of several empirical studies, we select the one which provides the richest insights on heterogeneity of ride-pooling behaviour. 
In the stated-preference survey on ride-pooling behaviour by \cite{alonso2021determinants}, the Latent Class Choice Model (LCCM) was estimated to fit the empirical results.
It indicates not only that people can be stratified into four distinct classes, but there is also a significant behavioural variability inside each class.

Travellers were revealed to be divided into classes defined as “It’s my ride” (C1, constituting $29\%$ of the sample), “Sharing is saving” (C2, $28\%$), “Time is gold” (C3, $24\%$), “Cheap and half empty, please” (C4, $19\%$). Participation in one of the classes is reflected in distinct behavioural parameters, namely the value of time (VoT) and penalty for sharing (PfS) varied among the classes. 

The structure of the utility formulas in \cite{alonso2021determinants} differed from the ExMAS utility formulas, thus, the following interpretations were made to apply the results to our study. 
Firstly, we assumed that both PfS and VoT in formula \ref{utility_new_exmas} follow a multimodal normal distribution given as a mixture of four normal distributions (corresponding to each class), like in eq. \ref{eq:random}.
The mean value of time is explicitly given in the study for respective classes as $16.98, 14.02, 26.25, 7.78$ (\texteuro /h) accordingly.
Yet, the penalty sharing needed to be derived from the reported willingness to share. We used PfS distribution reported for four additional passengers for class C1 as a reference (because such rides are admissible and appear in our experiments). Corresponding distributions for other groups were not reported, hence we retrieved them as scaled reference distribution. Applied proportions were following proportions of reported fixed WtS penalty for $2$ additional passengers.

Besides the mean values, we need to estimate their variances. Since this was not reported explicitly, we derive it from reported parameters estimates and z-values. 
To retrieve it, we apply the formula given by \cite{seltman2012approximations} assuming independent variables (covariance is null) and truncated normal distributions:
\begin{equation*}
    \textrm{Var}\left( \frac{X}{Y} \right) \approx \frac{\mu_x^2}{\mu_y^2} \left( \frac{\sigma_x^2}{\mu_x^2} + \frac{\sigma_y^2}{\mu_y^2} - 2 \frac{\textrm{Cov}(X, Y)}{\mu_x \mu_y} \right).
\end{equation*}
Which led to the variances for PfS of $0.082$, $0.071$, $0.06$ and $0.076$ and for VoT of $0.318$, $0.201$, $5.777$ and $13.063$ for the respective four classes. We clipped the huge variance for the last class to $1$, to prevent a class with practically random behaviour. Such big standard deviation presumably resulted from a small sample in a study.

Eventually, we successfully obtained: class membership probabilities ($p_{C}$), mean values of value-of-time and penalty for sharing for four different classes ($\beta_{s,C}$, $\beta_{t,C}$) and class-respective variances ($\sigma_{s,C}$, $\sigma_{t,C}$) as reported presented in Table \ref{tab:vot_PfS_values}.

\begin{table}[!h]
    \centering
    \resizebox{\textwidth}{!}{
    \begin{tabular}{cc|c||c|c|c|c}
          & & class & C1 & C2 & C3 & C4 \\ 
          & & name &  “It’s my & “Sharing is & “Time is & “Cheap and \\ 
        parameter  & &  &  ride” & “saving” & gold" & “ half empty" \\ \hline
         
        \multirow{2}{*}{VoT} & \multirow{2}{*}{$\beta_{t,C}$} & mean & 16.98 & 14.02 & 26.25 & 7.78 \\ 
        & & st.dev. & 0.318 & 0.201 & 5.777 & 1$^*$ \\ \hline
        \multirow{2}{*}{PfS} & \multirow{2}{*}{$\beta_{s,C}$} & mean  & 1.22 & 1.135 & 1.049 & 1.18 \\
        & & st.dev. &  0.082 & 0.071 & 0.06 & 0.076 \\ \hline
        Share & $p_C$ & & 29\% & 28\% & 24\% & 19\% \\ \hline
    \end{tabular}}
    \caption{\small Ride-pooling behavioural parameterization from \cite{alonso2021determinants}. Mean values and standard deviations of value of time (VoT) and penalty for sharing (PfS) for four classes. $^*$ for class 4 we arbitrarily clipped the otherwise too wide st. dev. of VoT }
    \label{tab:vot_PfS_values}
\end{table}

The total variance is distributed in formula \ref{utility_new_exmas} into a random value of $\beta$ and user-specific and ride-specific Gaussian noises. Since such distinction was not reported empirically, we assumed that standard deviation of the ride-specific noise $\epsilon_{i, r_k}$ (drawn for every evaluation of $U^s_{i,r_k}$) constitutes $10\%$ of standard deviation of user-specific noise (drawn once per replication) which follows standard normal distribution. Such noise is both consistent with empirical findings and mimics the additional uncertainty of the probit model (Cf. \cite{ortuzar2011}).

\subsection{Probabilistic shareability networks}
\begin{figure}
    \centering
        \begin{subfigure}[b]{0.32\textwidth}
        \centering
        \includegraphics[width=\textwidth]{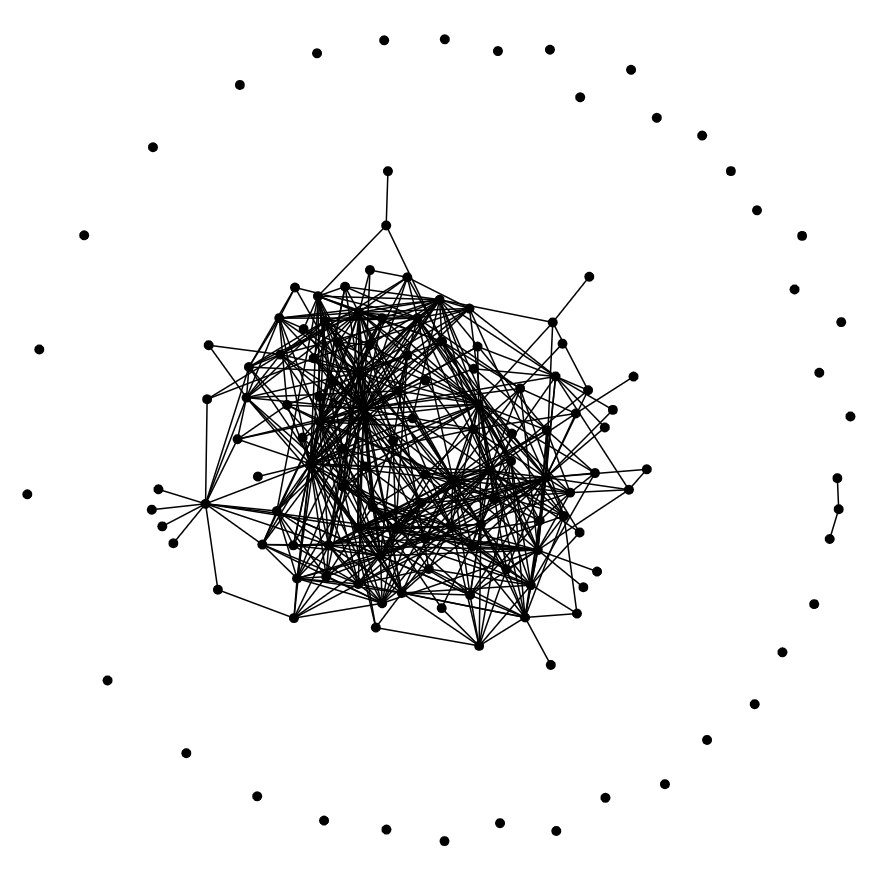}
        \caption{\small Shareability 1 rep.}
        \label{fig:single_pairs_shareability1}
    \end{subfigure}
    \hfill
        \begin{subfigure}[b]{0.32\textwidth}
        \centering
        \includegraphics[width=\textwidth]{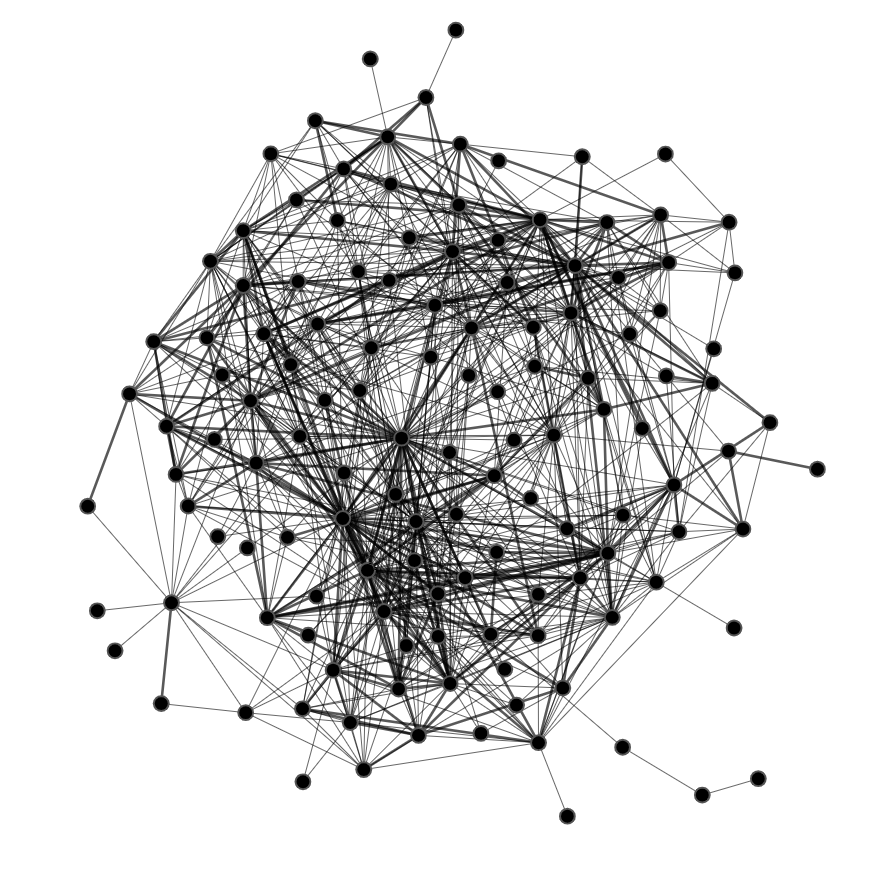}
        \caption{\small Shareability 2 rep.}
        \label{fig:single_pairs_shareability2}
    \end{subfigure}
    \hfill
        \begin{subfigure}[b]{0.32\textwidth}
        \centering
        \includegraphics[width=\textwidth]{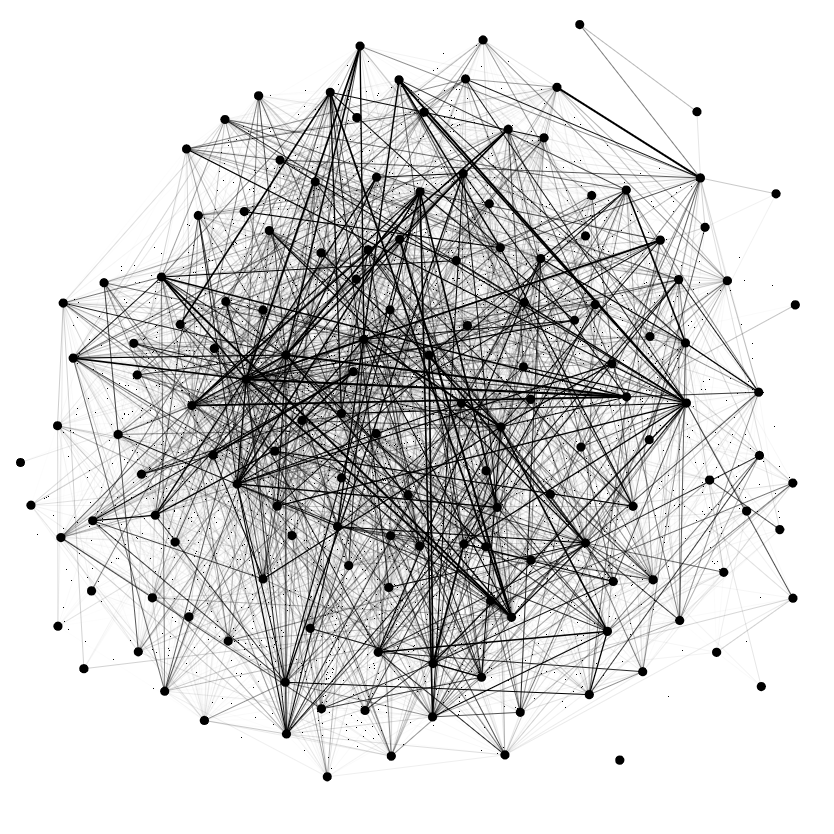}
        \caption{Shareability 1000 rep.}
        \label{fig:full_pairs_shareability}
    \end{subfigure}
    \vskip\baselineskip  
    \begin{subfigure}[b]{0.32\textwidth}  
        \centering 
        \includegraphics[width=\textwidth]{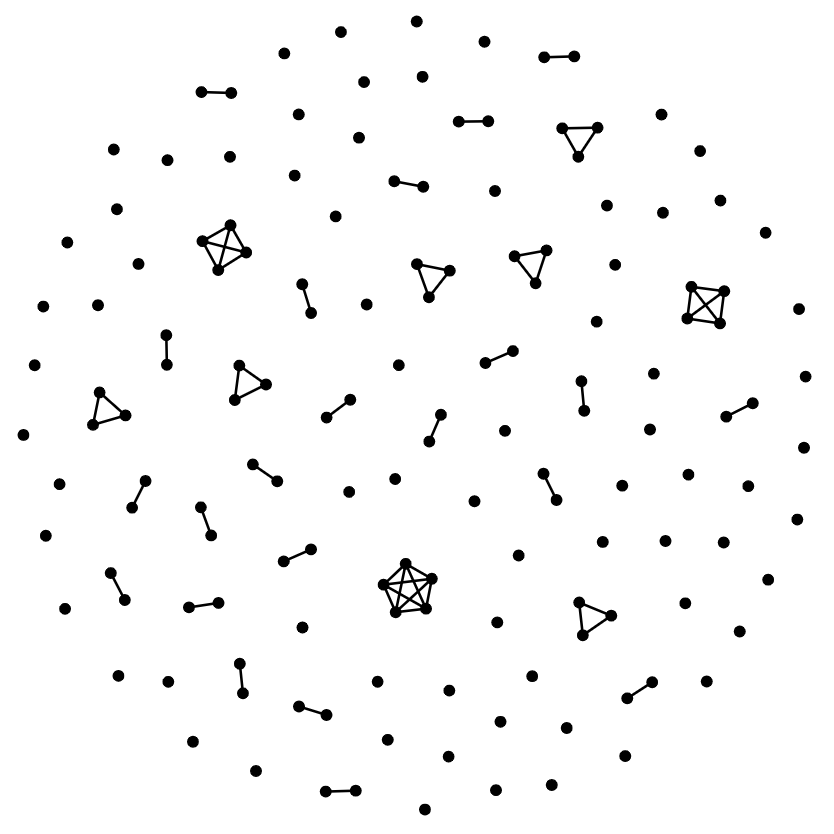}
        \caption{\small Matching 1 rep.}  
        \label{fig:single_pairs_matching1}
    \end{subfigure}
    \hfill
    \begin{subfigure}[b]{0.32\textwidth}  
        \centering 
        \includegraphics[width=\textwidth]{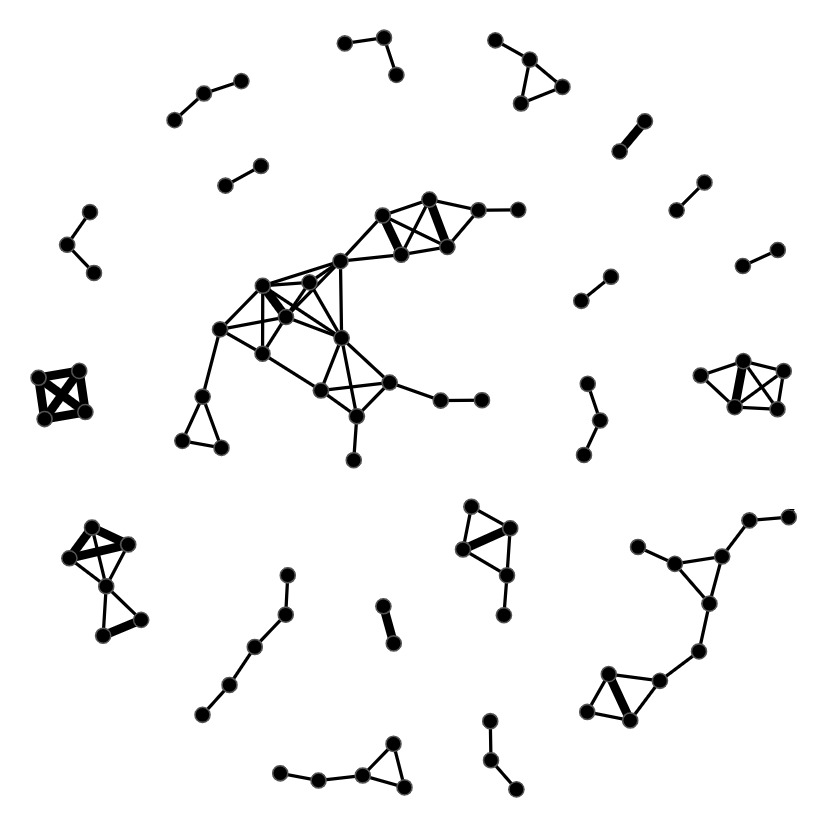}
        \caption{\small Matching 2 rep.}  
        \label{fig:single_pairs_matching2}
    \end{subfigure}
    \hfill
    \begin{subfigure}[b]{0.32\textwidth}  
        \centering 
        \includegraphics[width=\textwidth]{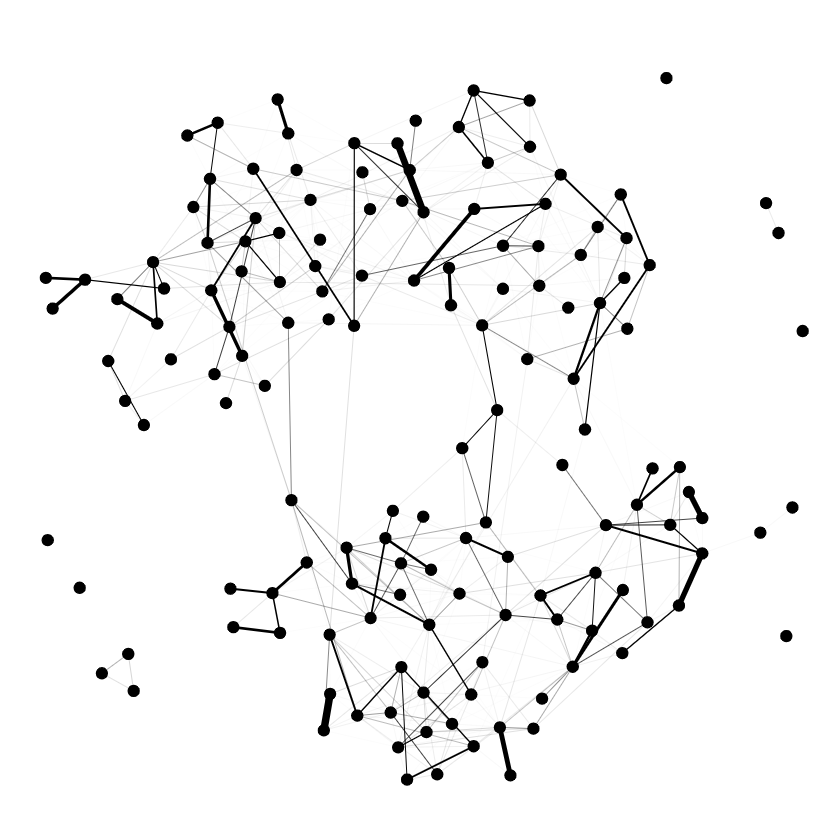}
        \caption{Matching 1000 rep.}
        \label{fig:full_pairs_matching}
    \end{subfigure}
    \caption{\small Shareability and matching networks obtained in the first, two and all replications of ride-pooling problem. Link widths (proportional to weights) denote how frequent two travellers were connected, i.e. found travelling together attractive (top row) or actually travelled together in the optimal solution (bottom row).}
    
    \label{fig:nets}
\end{figure}

To illustrate how the already complex shareability structures of ride-pooling change in the non-deterministic setting, we present them as the graph structures below. 
In Figure \ref{fig:nets}, we present shareability and matching structures resulting from first, first two and all replications of ride-pooling problem. 
In all networks, nodes represent each of $150$ travellers. Two travellers (nodes) are linked in the shareability graphs (\ref{fig:single_pairs_shareability1}, \ref{fig:single_pairs_shareability2}, \ref{fig:full_pairs_shareability}) if they can be attractively pooled together, i.e. formula \ref{utility_new_exmas}. In the matching networks (\ref{fig:single_pairs_matching1}, \ref{fig:single_pairs_matching2}, \ref{fig:full_pairs_matching}), links represent the final realisation of rides, i.e. solution of the bi-partite matching yielding minimal total mileage (see details in \cite{kucharski2020exmas}). Pair represents a ride comprising two travellers, a triangle - three, etc. While figures \ref{fig:single_pairs_shareability1}, \ref{fig:single_pairs_shareability2}, \ref{fig:full_pairs_shareability} reveal the algorithmic intermediate step of ExMAS, figures \ref{fig:single_pairs_matching1}, \ref{fig:single_pairs_matching2}, \ref{fig:full_pairs_matching} visualise an actual physical realisation of ride-pooling.

Already replicating the random process twice (second column in Fig. \ref{fig:nets}), reveals a high variability of resulting networks. Many links appear only in one replication and travellers either do not find travelling together attractive (in Fig. \ref{fig:single_pairs_shareability2}), or they are not part of the solution minimising the mileage (as on fig. \ref{fig:single_pairs_matching2}). This may be either due to their behaviour (high VoT and high penalty for sharing (PfS)) or due to their spatio-temporal incompatibility with others. The reason behind such variability lays in the assumption that we cannot predict the class memberships of individual travellers, thus an individual may be assumed to be a ride-pooling enthusiast (C2) in one replication and sharing-averse (C1) in another. While this is not true for individual travellers, this represents the perspective of the service provider who is not aware of the behaviour of a particular traveller.

This trend evolves, however, as we replicate the random process and observe diverse shareability and matching structures. 
We aggregate results from $1000$ replications and introduce the edges' weights (link widths) as linearly proportional to the number of occurrences among all replications. The weighted structures are presented in Figure \ref{fig:full_pairs_shareability} and Figure \ref{fig:full_pairs_matching}. Comparing figures \ref{fig:full_pairs_shareability} and \ref{fig:full_pairs_matching} to figures \ref{fig:single_pairs_shareability1} and \ref{fig:single_pairs_matching1}, we observe a significant increase in the number of links. In the probabilistic shareability network (Fig. \ref{fig:full_pairs_shareability}), only one traveller remained isolated after $1000$ replications (a quarter was isolated in one replication). The number of links increased nearly tenfold from about $600$ to $6000$. The matching network from a single replication  has $62$ links, while after $1000$ replications there are over $500$ links (Fig. \ref{fig:full_pairs_matching}).

Above results confirm that the ride-pooling problem for non-deterministic and heterogeneous travellers yield significantly different results. In the deterministic setting, the links can be well predicted as we can reasonably determine whether two travellers find it attractive to pool a ride. Yet, as soon as we include their varying behaviour more and more links are formed, and the structure becomes on one hand more dense and connected (as we aggregate along the replications) yet also more variable (when we compare two independent replications). It is substantially harder to predict whether two (or more) travellers will be happy to share a ride together when we know they may belong to classes of substantially different behaviours. That is why the aggregated shareability network becomes less interpretable as the number of feasible pairs significantly increases. 

On the other hand, some of the links are more frequently observed than the others. This is due to the fact that some trips are naturally shareable due to their spatio-temporal similarity (their origins and destinations are close to each other and the departure times are similar). Such similarity (or dissimilarity) can well compensate behavioural penalty for sharing and value of time. Disentangling the two factors (behaviour and spatio-temporal similarity) is not trivial, as they both drive the attractiveness of pooling. 

This initially chaotic representation becomes more rigid when we look at the matching solutions. 
Mind that in the matching, we optimise for the distance (which is deterministic), rather than user-varying random attractiveness.
This stabilises the shareability network and the link weights become more pronounced. Some pairs of travellers are connected with high probability, regardless of the varying behaviour of involved travellers. 
This demonstrates some regularity and predictability of ride-pooling, despite highly variable traveller behaviour, as we will demonstrate in more detail below.

\subsection{Expected ride-pooling performance with non-deterministic travellers}\label{sec:performance_assesment}
\begin{figure}
    \centering
    \begin{subfigure}[b]{0.243\textwidth}   
        \centering 
        \includegraphics[width=\textwidth]{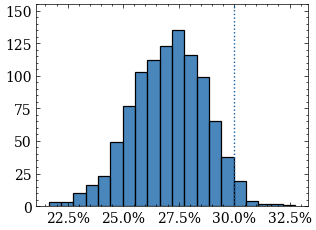}
        \caption{\footnotesize Mileage reduct. $\mathcal{D}$}
        \label{fig:vehhours147}
    \end{subfigure}
    \hfill
        \begin{subfigure}[b]{0.22\textwidth}  
        \centering 
        \includegraphics[width=\textwidth]{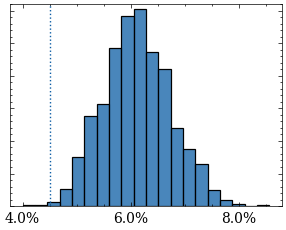}
        \caption{\footnotesize Utility gain $\mathcal{U}$}
        \label{fig:passutility147}
    \end{subfigure}
    \hfill
    \begin{subfigure}[b]{0.22\textwidth}
        \centering
        \includegraphics[width=\textwidth]{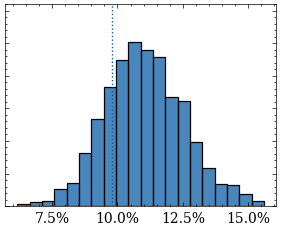}
        \caption{\footnotesize Trip detour $\mathcal{T}$}
        \label{fig:passhours147}
    \end{subfigure}
    \hfill
    \begin{subfigure}[b]{0.22\textwidth}   
        \centering 
        \includegraphics[width=\textwidth]{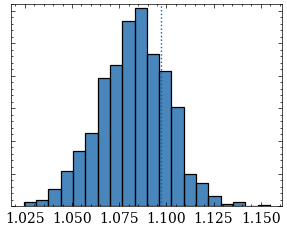}
        \caption{\footnotesize Profitability $\mathcal{P}$}
        \label{fig:profitability147}
    \end{subfigure}
    \caption{\small Distribution of ride-pooling performance resulting from 1000 replications. Dotted lines show the deterministic benchmark.}
    \label{fig:kpis}
\end{figure}

In this section, we dive into details how the behavioural heteroscedasticity impacts the ride-pooling performance. We assess the ride-pooling system’s performance with the four indicators introduced in Section \ref{sec:performance_indicators}. Calculated first for a consecutive replications (realisations of random variables) and then accumulated over all 1000 replications (one value corresponds to one realisation). Results are presented in Figure \ref{fig:kpis}. Baseline for comparison is the deterministic ExMAS (with VoT and PfS being weighted average of means introduced in Table \ref{tab:vot_PfS_values}) which yields mileage reduced by $30\%$,  $9.8\%$ detour, utility increased by $4.5\%$ and profitability of $1.097$. 

Regardless of behavioural variability, pooling always reduces the travel distance ($\mathcal{D}$). $90\%$ of the observations range between $24.3\%$ and $29.7\%$ with the average of $27.1\%$, which is significantly worse than 30\% reductions obtained in the deterministic setting (Figure \ref{fig:vehhours147}). While this is a significant reduction, which should be hugely appreciated from the city’s and environmental’s perspective, it also varies significantly, with observations ranging from $20\%$ to $32.5\%$. The right tail seems to be fatter, with occasional outliers greater than $30\%$ vehicle kilometres reductions.
The mean of average utility gain (which is user-subjective and depends on behavioural profile) is approximately $6.1\%$, which, in turn, is now greater than in the deterministic benchmark. The $90\%$ two-sided confidence interval spans between $5.1\%$ and $7.2\%$ utility gains.

A more physical measure of pooling performance, from the user’s perspective, is trip detour, which does not depend on traveller behaviour. Notably, the acceptable detour (and delay), unlike the fixed time windows used in other studies, depends on VoT and PfS and can vary significantly with the behaviour. 

The mean travel time of pooled ride is approximately $11\%$ longer than for the solo-rides only system (Figure \ref{fig:passhours147}) and $90\%$ of the observations fit between $8.6\%$ and $13.6\%$, which, again, is worse than in the deterministic benchmark ($9.8\%$).

Regardless of the behavioural variability, the ride-pooling remains profitable, with average profitability of $1.082$ and the $90\%$ confidence interval from $1.05$ to $1.11$ (Fig. \ref{fig:profitability147}). It is a very promising finding. It means that in $95\%$ of scenarios, profitability gain was greater than $5\%$ and we did not observe any unprofitable replication, meaning that the platform does not risk when offering a 30\% discount for pooled rides.

\subsection{Class dependent ride-pooling performance}\label{sec:performance_classes}

Considering the importance of the behavioural preferences of individuals on pooling performance, we analyse further the data with respect to individuals and their classes.
First, we show the cumulative distributions of individual indicators observed in $1000$ replications. Along with overall distributions, we present the profiles associated with four distinct classes. While the CDF profiles serve for illustrative purposes mainly, we enhance them with tables showing differences between classes in the four consecutive indicators as they reveal intriguing patterns among the introduced behavioural classes.    

\begin{figure}
    \centering
        \begin{subfigure}[b]{0.45\textwidth}  
        \centering 
        \includegraphics[width=\textwidth]{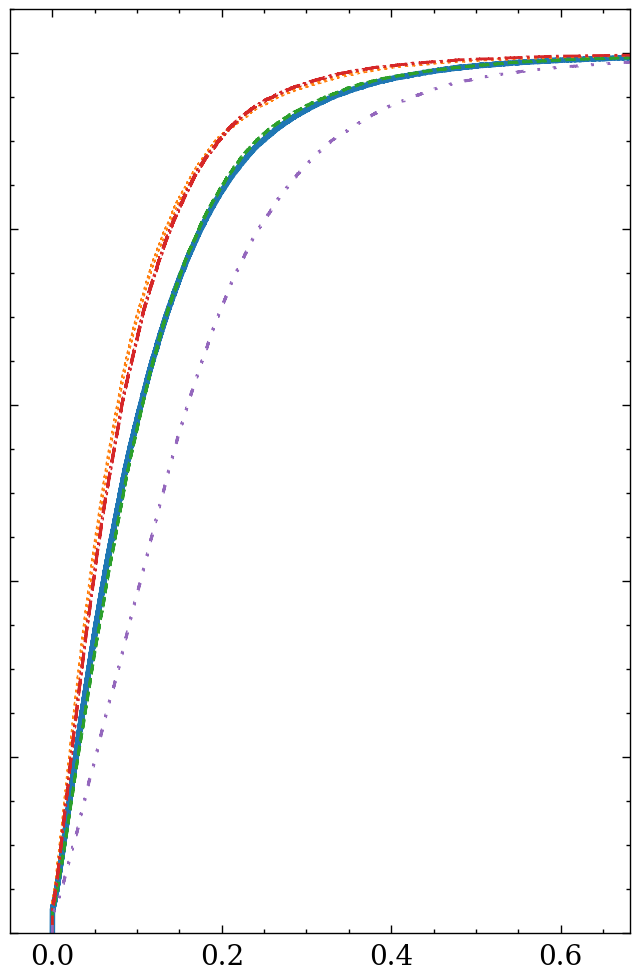}
        \caption{\small Utility gain ($\mathcal{U}_r$)}
        \label{fig:cdf_u}
    \end{subfigure}
    \hfill
    \begin{subfigure}[b]{0.45\textwidth}
        \centering
        \includegraphics[width=\textwidth]{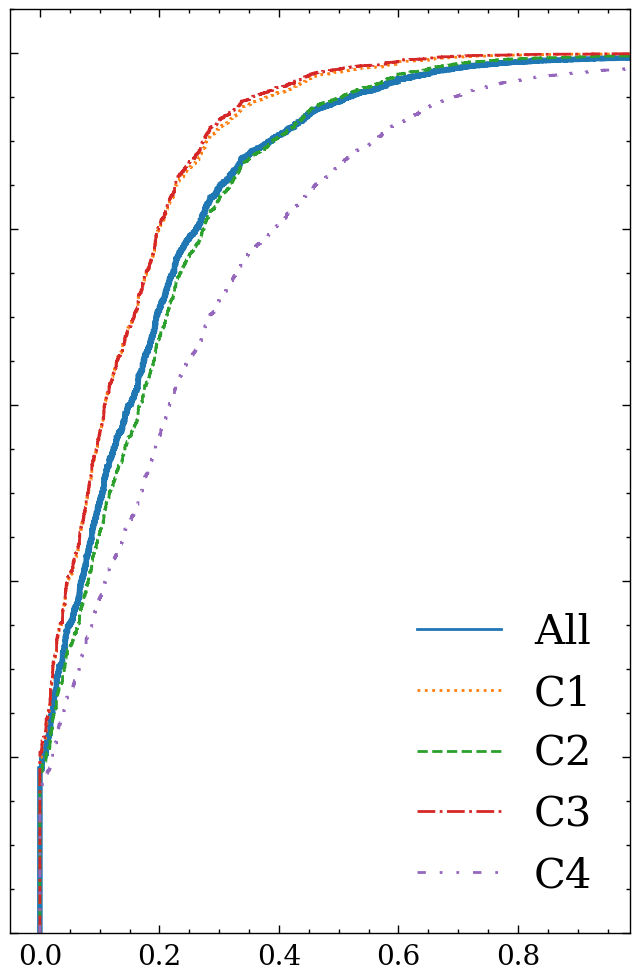}
        \caption{\small Detour ($\mathcal{T}_r$)}
        \label{fig:cdf_pass}
    \end{subfigure}
    \caption{\small Cumulative distribution functions of utility gains and detours for the trips observed in 1000 replications, stratified into four classes of behaviour. }
    \label{fig:cdfs}
\end{figure}

Here, we aggregate all the rides, irrespective of in which replication they appear.
In Fig. \ref{fig:cdfs}, we present the cumulative distribution functions of the utility gains and detours. 
For classes C1 and C3, the utility gains are below average and detours are above average. For the class C4 the opposite is true, while class C2 in on average with the mean performance.

In Table \ref{tab:delay_all}, we report the resulting trip detour for the passengers in the four respective behavioural classes classes.
Apart from mean and standard deviation, we report values of $75$th, $90$th percentile and $95$th percentile. 
First, we present the values obtained for all the rides and then, for illustrative purposes, only for shared rides (excluding the travellers who were not matched with anyone and travelled alone). While on average, pooled rides are 8\% longer than private rides, it varies from 5\% for class C1 to 15\% for class C4. Notably, 5\% of travellers in C4 decided to pool despite at least 61\% longer travel time. Those trends are further pronounced when we restrict the analysis to the rides which were successfully pooled (right columns).

\begin{table}[!ht]
    \centering
    \resizebox{\textwidth}{!}{
    \begin{tabular}{c|c|c|c|c|c||c|c|c|c|c}
    
    {} & \multicolumn{5}{c||}{All rides} & \multicolumn{5}{c}{Shared rides} \\
    \hline
    {} &                        Means &  St.dev. &   75 &   90 &   95 &         Means &  St.dev. &   75 &   90 &   95\\
    \hline
    All &                       0.08 &    0.16 & 0.11 & 0.27 & 0.39 &           0.16 &    0.19 & 0.22 & 0.39 & 0.52 \\
    C1 “It’s my ride” &         0.05 &    0.11 & 0.07 & 0.19 & 0.27 &           0.12 &    0.13 & 0.18 & 0.28 & 0.38 \\
    C2 “Sharing is saving” &    0.09 &    0.15 & 0.14 & 0.28 & 0.41 &           0.16 &    0.17 & 0.23 & 0.39 & 0.50 \\
    C3 “Time is gold” &         0.05 &    0.11 & 0.07 & 0.19 & 0.27 &           0.11 &    0.13 & 0.18 & 0.27 & 0.35 \\
    C4 “Cheap and half empty" & 0.15 &    0.23 & 0.23 & 0.45 & 0.61 &           0.23 &    0.25 & 0.33 & 0.56 & 0.69 \\
    
    \end{tabular}}
    \caption{\small Statistical properties of trip detour (relative) $\mathcal{T}_r$ for travellers of four respective behavioural classes. Mean, variance and significant percentiles are shown first for all the travellers (left) and then (right) only for those who were pooled.}
    \label{tab:delay_all}
\end{table}

Similarly to travel time analysis in Table \ref{tab:delay_all}, we present data of relative utility gain in ride-hailing system with ride-pooling (relative to system without pooling).
Surprisingly, the class which has the longest detours (C4) has also the greatest utility gains (11\% on average). For the class unwilling to pool the benefits are only 4\% (C1). For 10\% of all the travellers ride-pooling led to at least 25\% increase in perceived utility.

\begin{table}[!ht]
    \centering
    \resizebox{\textwidth}{!}{
    \begin{tabular}{c|c|c|c|c|c||c|c|c|c|c}
    
    {} & \multicolumn{5}{c||}{All rides} & \multicolumn{5}{c}{Shared rides} \\
    \hline
    {} &  Means &  St.dev. &   75 &   90 &   95 & Means &  St.dev. &   75 &   90 &   95\\
    \hline
    All &   0.06 &    0.10 & 0.09 & 0.18 & 0.25 & 0.11 &    0.12 & 0.15 & 0.25 & 0.33 \\
    C1 “It’s my ride” &   0.04 &    0.08 & 0.05 & 0.12 & 0.18 & 0.09 &    0.10 & 0.12 & 0.19 & 0.26 \\
    C2 “Sharing is saving” &   0.06 &    0.10 & 0.09 & 0.18 & 0.25 & 0.11 &    0.12 & 0.15 & 0.24 & 0.32 \\
    C3 “Time is gold” &   0.04 &    0.08 & 0.06 & 0.14 & 0.19 & 0.09 &    0.10 & 0.12 & 0.20 & 0.26 \\
    C4 “Cheap and half empty" &   0.11 &    0.14 & 0.16 & 0.28 & 0.37 & 0.16 &    0.15 & 0.22 & 0.33 & 0.42 \\
    
    \end{tabular}}
    \caption{\small Statistical properties of travellers gains (in relative increase of utility) $\mathcal{U}$ for travellers of four respective behavioural classes. Mean, variance and significant percentiles are shown first for all the travellers (left) and then (right) only for those who were pooling.}
    \label{tab:utility_all}
\end{table}

\subsection{Asymmetrical impact of perceived sharing discomfort}

Importantly, ride-pooling behaviour of an individual traveller may have diverse impact on system performance. While travellers unwilling to share form isolated nodes in the shareability graph, the travellers with high willingness to share become super connected and can share with many other travellers and join multiple groups. Thus, the importance of highly shareable travellers is way more important for the ride-pooling performance than single isolated solo-riders. 

\begin{figure}
    \centering
    \includegraphics[width=.65\textwidth]{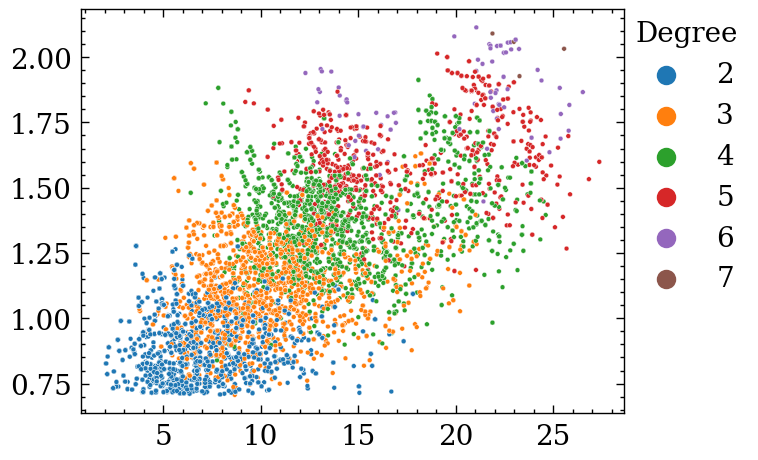}
    \caption{\small Relation between shared ride's distance (x-axis, in kilometres) and profitability (y-axis). Each dot denotes a ride observed in one of 1000 repliactions and is coloured accordingly to is degree (number of travellers).}
    \label{fig:rides_profitabilit}
\end{figure}

We illustrate impact of this this asymmetry with Figure \ref{fig:rides_profitabilit}, showing a relation between the distance of the shared ride and its profitability. Colours show the degree of a ride, i.e. number of travellers in the ride. In the deterministic benchmark, the maximal observed degree was five, however in the probabilistic experiment, rides composed of six and seven travellers were observed. This is thanks to the asymmetrical impact of highly shareable travellers on the shareability networks. Which is profound also for the ride-profitability, as the strong positive trend between the ride degree and profitability is observable in Fig. \ref{fig:rides_profitabilit}.

\subsection{Ride-pooling performance vs ride-pooling behaviour}
\begin{figure}[!h]
    \centering
    \begin{subfigure}[b]{0.48\textwidth}
        \centering
        \includegraphics[width=\textwidth]{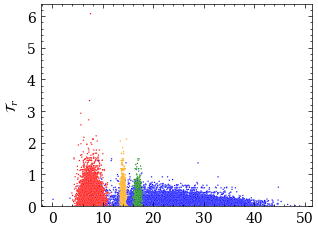}
        \caption{\small $\mathcal{T}_r$ as a function of VoT}
        \label{fig:scatter_vot_time}
    \end{subfigure}
    \hfill
    \begin{subfigure}[b]{0.48\textwidth}  
        \centering 
        \includegraphics[width=\textwidth]{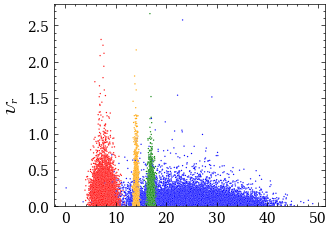}
        \caption{\small $\mathcal{U}_r$ as a function of VoT}
        \label{fig:scatter_vot_util}
    \end{subfigure}
    \vskip\baselineskip
    \begin{subfigure}[b]{0.48\textwidth}   
        \centering 
        \includegraphics[width=\textwidth]{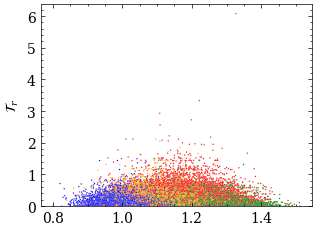}
        \caption{\small $\mathcal{T}_r$ as a function of PfS}
        \label{fig:scatter_wts_time}
    \end{subfigure}
    \hfill
    \begin{subfigure}[b]{0.48\textwidth}   
        \centering 
        \includegraphics[width=\textwidth]{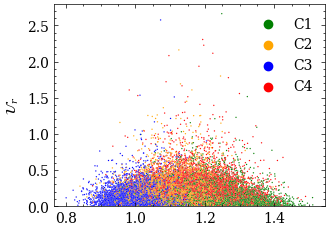}
        \caption{\small $\mathcal{U}_r$ as a function of PfS}
        \label{fig:scatter_wts_util}
    \end{subfigure}
    \caption{\small Scatter plots of travellers' behavioural parameters ($x$-axis) against the resulting ride-pooling performance. Each dot denotes a single traveller, coloured accordingly to the class assigned to her in the respective replication. y-axis are scaled to represent relative change, e.g. $1$ corresponds to increase of $100\%$.}
    \label{fig:scatters}
\end{figure}

In this section, we provide an insight into the relation between sampled behavioural parameters (VoT, PfS) and performance of the indivudal travellers ($\mathcal{U}_{r}$, $\mathcal{T}_{r}$).
In Figure \ref{fig:scatters}, we scatter the values of VoT and PfS (first and second row respectively) on $x$-axis against the performance indicators, i.e. $\mathcal{T}_r$ and $\mathcal{U}_r$ (first and second column) on $y$-axis. Each dot represent an individual traveller coloured accordingly to her behavioural class.

We can observe a clear trend of ride-pooling detours decreasing with an increasing value of time (Fig. \ref{fig:scatter_vot_time}), despite the longer detours for travellers with low value of time, their benefits of pooling (relative increase in utility) remains high (Fig. \ref{fig:scatter_vot_util}), whereas for travellers with high value of time utility gains are significantly lower. The three classes (C1, C2 and C4) have relatively low variance and do not overlap, while for class C3 the variance of VoT is very big. Mind that in our experiment each traveller was assigned to different classes across replications, nonetheless the class membership strongly correlate with the resulting ride-pooling performance, despite the fixed spatiotemporal trip characteristics.

Those clear trends become blurred when we plot against PfS (Fig. \ref{fig:scatter_wts_time} and \ref{fig:scatter_wts_util}) where the Gaussian shape is observed yet class memberships are indistinguishable. Both highest detours and utility gains are obtained for intermediate values of PfS, while extreme cases have the lowest benefits. The travellers with a low penalty for sharing (below 1 for a subset of class C3) seem to be often exploited by the system and their utilities are sacrifices to match preferences for travellers with lower preferences for sharing. Yet for those with high penalty the benefits are also limited, since they can hardly be pooled with others. The pooling seems to be best performing with the penalties ranging from 1.1 to 1.2 where it can reach the greatest expected benefits (Fig. \ref{fig:scatter_wts_util}).

\subsection{Scaling with the demand levels}
The notion of critical mass is crucial to the ride-pooling and number of studies reported how the ride-pooling grows non-linearly with the demand size. Yet, up to now, those findings were reported in the deterministic setting only. Here, apart from reporting how system performance improves with growing demand, we provide insights on how the performance variability changes.

\begin{figure}
    \centering
    \begin{subfigure}[b]{0.23\textwidth}   
        \centering 
        \includegraphics[width=\textwidth]{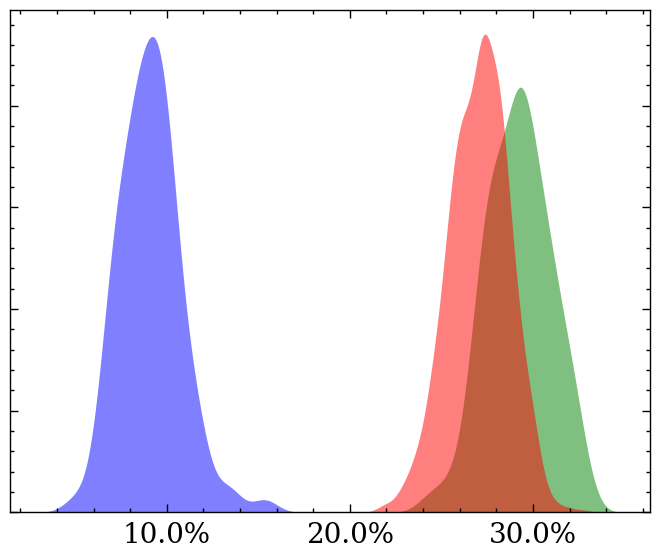}
        \caption{\footnotesize Mileage save ($\mathcal{D}$)}
        \label{fig:mixedvehhours}
    \end{subfigure}
    \hfill
        \begin{subfigure}[b]{0.23\textwidth}  
        \centering 
        \includegraphics[width=\textwidth]{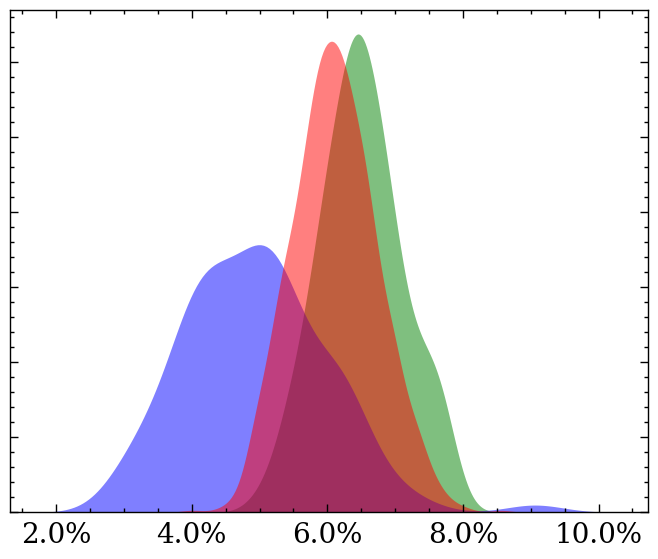}
        \caption{\footnotesize Utility gain ($\mathcal{U}$)}
        \label{fig:mixedutility}
    \end{subfigure}
    \hfill
    \begin{subfigure}[b]{0.23\textwidth}
        \centering
        \includegraphics[width=\textwidth]{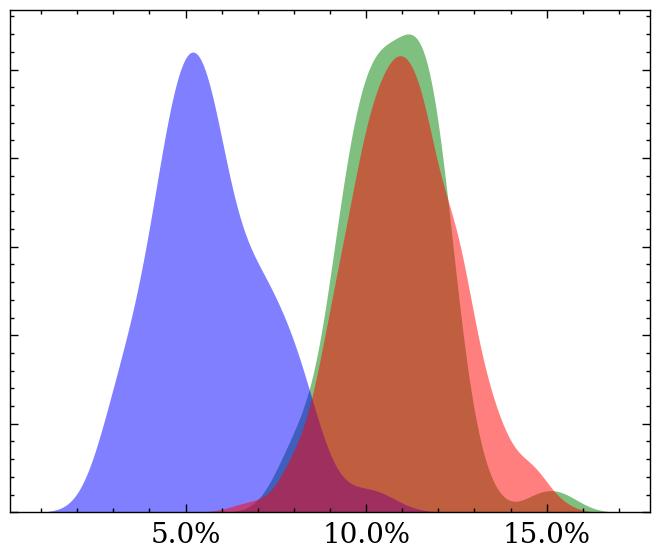}
        \caption{\footnotesize Trip detour ($\mathcal{T}$)}
        \label{fig:mixedpasshours}
    \end{subfigure}
    \hfill
    \begin{subfigure}[b]{0.23\textwidth}   
        \centering 
        \includegraphics[width=\textwidth]{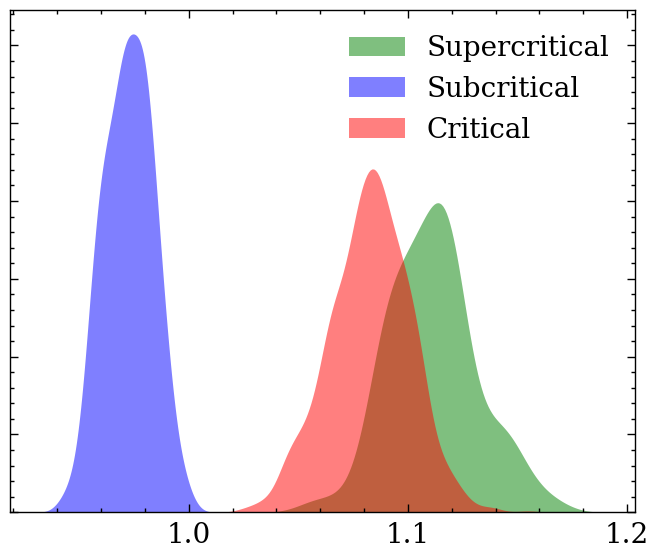}
        \caption{\footnotesize Profitability ($\mathcal{P}$)}
        \label{fig:mixedprofitability}
    \end{subfigure}
    \caption{\small Distributions of four performance indicators with the three levels of demand: subcritical, critical and supercritical.}
    \label{fig:kpis_small_big}
\end{figure}

Demand set on which we run the experimental was selected to be slightly beyond the critical mass needed to induce the effective pooling, we extend it here from two sides: sub-critical ($99$ requests) and super-critical ($198$ requests). We report the present previously introduced performance measures for three levels of demand in Figure \ref{fig:kpis_small_big}.  
Unsurprisingly, we observe the critical mass effect as the performance significantly increases when the demand grows from 100 to 150 requests and then stabilizes when it reaches 200 requests (Fig. \ref{fig:scaling}).
Yet more importantly, we observe the trends with variabilities, which are alternative: either narrowing with growing demand (utility in Fig. \ref{fig:mixedutility}), widening (profitability in Fig. \ref{fig:mixedutility}), or remaining roughly constant (travel times and mileage savings in Fig. \ref{fig:mixedvehhours} and Fig. \ref{fig:mixedpasshours}).

 \label{fig:scaling}

\section{Discussion and conclusions}
Most ride-pooling algorithms rely on the fixed constraints (usually time windows). To more objectively assess the attractiveness of a shared ride from the traveller’s perspective, we previously proposed the ExMAS algorithm. In this study, we take the matter further. We no longer consider travellers homogeneous and deterministic. Using recent empirical findings, we assume that while we still do now know preferences of the individuals, we know the general distributions.

To properly include this in our calculations, we introduced random variables representing the value of time and perceived sharing discomfort. Such reformulation allowed to reproduce the four distinct classes of travellers' behaviour towards ride-pooling, as revealed in \cite{alonso2021determinants}. Conducting sufficient number of replications, we obtain reliable estimations of the ride-pooling problem solution, which is no longer deterministic. Apart from mean values, we can estimate the lower and upper bounds of the performance indicators, which substantially differ from the deterministic benchmark. 

The additional probabilistic layer introduces a variability at the level of individuals and at the system performance. We find that despite the high variability of individuals, pooling performance remains stable and within reasonably narrow confidence intervals. Notably, probabilistic results are shifted compared to the deterministic benchmark. While the primary objective of minimising mileage is better met in the deterministic scenario, we observe a much higher satisfaction with the service in the heterogeneous setting.

Our study also provides an insight into how ride-pooling performs for travellers of certain preferences. We analyse the impact of the value of time and perceived sharing discomfort on trip detour and satisfaction with the service. We find that people with low value of time can be considered both the most flexible and the most beneficial travellers in the pooling system. However, those with intermediate penalty for sharing not only benefit more than those with high penalties (who does not want to share in general), yet also more than those with low penalties (willing to share with anyone).

Similarly to the deterministic case, we observe the critical mass effect and pooling becomes effective only when the demand levels reach the so-called critical mass. Now we enrich this notion with findings on variability, which may either decrease (in terms of utility gains) or increase (in terms of profitability for the provider) with growing demand levels.

The proposed method is general and can be easily applied to new cases, both for general demand patterns and different behavioural models. Also, the specific experimental setting used in this study may be reformulated, e.g. when we know the class membership, the demand is not predicted properly or when the behaviour is assumed fixed, but demand is varying (like in \cite{kucharski2021virus}). In the future studies, those additional dimensions of variability may be included for even richer assessments.

Finally, the proposed method may be valuable in the pandemic-analyses, when virus-averse behaviour drives the pooling behaviour of individuals. With the methodology that we propose, one may better understand the performance of ride-pooling system under dynamically changing behaviour, which becomes highly variable during disruptive changes. This can be instrumental in the event of the new pandemic. One can now assess the potential of ride-pooling and balance the performance with the spreading contribution. Notably, the COVID-19 risk-awareness was heterogeneous and non-deterministic in the population. We argue that the penalty-for-sharing, with which we experiment in this study, may be interpreted as the virus-awareness and assess with the proposed method.

\paragraph{Acknolwedgements}
This research was funded by National Science Centre in Poland program OPUS 19 (Grant Number 2020/37/B/HS4/01847)

\bibliographystyle{cas-model2-names}

\bibliography{cas-refs}





\end{document}